\documentclass{ws-ijgmmp}

\begin{document}

\markboth{G Marmo, G Scolarici, A Simoni, F Ventriglia } { The
Quantum-Classical Transition: the Fate of the Complex Structure }

%
\catchline{}{}{}{}{}
%

\title{THE QUANTUM-CLASSICAL TRANSITION: THE FATE OF THE COMPLEX STRUCTURE }

\author{G. MARMO\footnote{Dip. Sc. Fisiche, Universit\'a Federico II,
Via Cintia 80126 Napoli, Italy.}}

\address{Dip. Sc. Fisiche and INFN-Napoli, Universit\'a Federico II, Via Cintia \\
Napoli, 80125, Italy
\\
\email{marmo@na.infn.it}}

\author{G. Scolarici}

\address{Dip. Fisica and INFN-Lecce, Universit\'a di Lecce, Via per Arnesano\\
Lecce, 73100, Italy
\\
\email{scolarici@le.infn.it}}

\author{A. Simoni}

\address{Dip. Sc. Fisiche and INFN-Napoli, Universit\'a Federico II, Via Cintia \\
Napoli, 80125, Italy
\\
\email{simoni@na.infn.it}}

\author{F. Ventriglia}

\address{Dip. Sc. Fisiche and INFN-Napoli, Universit\'a Federico II, Via Cintia \\
Napoli, 80125, Italy
\\
\email{ventriglia@na.infn.it}}

\maketitle

\begin{history}
\received{(Day Month Year)}
\revised{(Day Month Year)}
\end{history}

\begin{abstract}
According to Dirac, fundamental laws of Classical Mechanics should
be recovered by means of an "appropriate limit" of Quantum
Mechanics. In the same spirit it is reasonable to enquire about the
fundamental geometric structures of Classical Mechanics which will
survive the appropriate limit of Quantum Mechanics. This is the case
for the symplectic structure. On the contrary, such geometric
structures as the metric tensor and the complex structure, which are
necessary for the formulation of the Quantum theory, may not survive
the Classical limit, being not relevant in the Classical theory.
Here we discuss the Classical limit of those geometric structures
mainly in the Ehrenfest and Heisenberg pictures, i.e.  at the level
of observables rather than at the level of states. A brief
discussion of the fate of the complex structure in the
Quantum-Classical transition in the Schroedinger picture is also
mentioned.
\end{abstract}


\section{Introduction}

According to Dirac \cite{1}:

\textsl{``Classical mechanics must be a limiting case of quantum
mechanics. We should thus expect to find that important concepts in
classical mechanics correspond to important concepts in quantum
mechanics and, from an understanding of the general nature of the
analogy between classical and quantum mechanics, we may hope to get
laws and theorems in quantum mechanics appearing as simple
generalizations of well known results in classical mechanics''.}

In the Dirac approach to Quantum Mechanics, the carrier space is a
Hilbert space, i.e. it is a vector space endowed with a Hermitian
Structure which give rise to a Riemannian Structure (the real part
of the inner product), a symplectic structure (the imaginary part of
the inner product) and a connecting (1-1)-tensor, the complex
structure $J$ satisfying $J^{2}=-1$.

On the other hand in Classical Mechanics the prevailing structure is
provided by the symplectic structure, it is therefore a natural
question to ask what happens, in the appropriate limit as suggested
by Dirac, of both the complex and Riemannian structures.

Of course a similar question can be asked with respect to the linear
structure. Here, however, the situation is more involved because,
due to the probabilistic interpretation of Quantum Mechanics, one
requires that only the probability density is physically meaningful
and not the probability amplitude, this means that from the
measurement view point, the carrier space of Quantum Mechanics is
the complex projective space associated with $\Bbb{H} $ rather than
$\Bbb{H}$ itself. On this manifold there is still a K\"{a}hler
structure \cite{Weil} but the linear structure has disappeared ,
showing that it is not the linear structure which matters but rather
a superposition (composition rule) for solutions, allowing to
construct new solutions out of two given ones, as it is well known
this phenomenon is well understood in the framework of Riccati-type
equations and we will not insist any further on the subject, we
refer to the existing literature \cite{Carinena} on the argument.

As for Quantum Mechanics, there has been a recent proposal to define
a superposition rule on pure states \cite{Sudarshan} (the complex
projective space of $\Bbb{H}$) which amounts to the introduction of
a connection on the complex projective space. This connection is
essentially given by the Pancharatnam phase, which is also known in
the literature as the Berry phase or geometric phase.\cite
{PanchaMukunda}

Apart from the details concerning the linear structure or the
superposition principle of Quantum Mechanics, the K\"{a}hler
structure is always present and it is a challenge to understand what
happens of the complex structure when the appropriate limit to go
from Quantum to Classical is taken. Because we want to consider the
fate of the complex structure in the Quantum-Classical transition,
it is more appropriate to deal directly with a real vector space and
a complex structure $J$ defined on it, rather than consider the
Hilbert space and hide the complex structure in the defining
properties of the space itself.

Before entering our discussion we should point out that the role
played by the complex structure in Quantum Mechanics was already
considered by H. Reichenbach in his book ''\textit{Philosophic
Foundations of Quantum Mechanics'' }.\cite{Reichenbach} It was also
extensively considered by Stueckelberg and his
collaborators.\cite{Stuec} Afterwards we find comments by G.W.
Mackey \cite {Mackey} and a more extensive treatment by V. Cantoni.
\cite {Cantoni}

A decade ago a new analysis started \cite{Cirelli} also in
connection with a proposal by S. Weinberg \cite{Weinberg} on a
non-linear Quantum Mechanics.

In referring to previous papers we have not distinguished between
papers actually focused on the role of the complex structure and
those dealing with the role of the Riemannian structure, the reason
being that for us the symplectic structure is not questioned because
(as a structure) it survives the ''appropriate limit'' to Classical
Mechanics. Therefore, with a symplectic structure granted, the
complex structure and the Riemannian structure are not independent
and each one determines the other one.

Because we are mostly interested in the Classical limit, we are
obliged to take into account various pictures of Quantum Mechanics
to consider the Classical limit separately for each one of them.
Traditionally, one considers the Schroedinger picture and the
Heisenberg picture, in which the equations of motion are considered
on the space of states or on the algebra of observables
respectively. Here we would like to replace the Heisenberg picture
with what we may call the Ehrenfest picture, that is we deal with
quadratic functions, expectations values, rather than with
operators.

\section{ Indetermination relations: the need of a complex structure}

In this section we discuss in a slightly generalized form the quoted
argument of Stueckelberg which shows the necessity of introducing a
complex structure $J$ to get indetermination relations and formulate
Quantum Mechanics. Therefore we start with a real Hilbert space .

An indetermination relation \emph{\`{a} la} Heisenberg usually is a
relation like
\begin{equation}
\left\langle \Delta A^{2}\right\rangle _{\Psi }\left\langle \Delta
B^{2}\right\rangle _{\Psi }\geq \left\langle i[A,B]\right\rangle
_{\Psi }^{2}
\end{equation}
where the error operators of the \textsl{observables } $A$ and $B$%
\begin{equation}
\Delta A=A-\left\langle A\right\rangle _{\Psi };\Delta
B=B-\left\langle B\right\rangle _{\Psi },
\end{equation}
are related to their commutator through the Hermitian operator
$i[A,B]$.

However, on a \textsl{real} Hilbert space, the multiplication by the
imaginary unit $i$ in $i[A,B]$ does not make sense. Moreover,
$[A,B]$ is a
skew-Hermitian operator, so that its mean value with respect to a \textsl{%
real} scalar product always vanishes.

Then, motivated by scaling arguments, in principle there are two
possibilities :
\begin{equation}
\left\langle \Delta A^{2}\right\rangle _{\Psi }\left\langle \Delta
B^{2}\right\rangle _{\Psi }\geq \left\{
\begin{array}{c}
\mu ^{2}\left\langle -[A,B]^{2}\right\rangle _{\Psi } \\
\mu ^{2}\left\langle K(A,B)\right\rangle _{\Psi }^{2}
\end{array}
\right.
\end{equation}
where $\mu $ is a real number and $K(A,B)$ a Hermitian bilinear
expression to be determined .

The first possibility leads to a contradiction for bounded operators
unless $\mu =0:$

In fact, choosing $\Psi $ in an orthonormal basis \{$\Phi _{n}$\} in
which $A$
is diagonal: $\left\langle \Delta A^{2}\right\rangle _{\Psi }=0$ and $%
\left\langle \Delta B^{2}\right\rangle _{\Psi }$ is bounded if $B$
is bounded, so their product vanishes while on the contrary
$\left\langle -[A,B]^{2}\right\rangle _{\Psi
}=\sum\nolimits_{n}(A_{\Psi }-A_{n})^{2}\left\langle \Psi |B|\Phi
_{n}\right\rangle ^{2}$ is strictly positive.

Only the existence of a complex structure $J$ leads to an
indetermination relation:

An indetermination relation (inequality) holds on a real Hilbert
space for Hermitian operators $A$,$B$ (observables) commuting with a
complex structure $J$. It has the form:

\begin{equation}
\left\langle \Delta A^{2}\right\rangle _{\Psi }\left\langle \Delta
B^{2}\right\rangle _{\Psi }\geq \frac{1}{4}\left\langle
J[A,B]\right\rangle _{\Psi }^{2}
\end{equation}

The following is a detailed proof which shows that a generalization
is possible. Consider the following real scalar product:
\begin{equation}
\left\langle (\Delta A+\lambda X\Delta B)\Psi ,(\Delta A+\lambda
X\Delta B)\Psi \right\rangle \geq 0,
\end{equation}
where $\lambda $ is a real number and $X$ is an operator which has
to be determined in such a way that an indetermination relation
holds. Expanding the preceding expression leads to
\begin{equation}
\left\langle \Delta A^{2}\right\rangle _{\Psi }+\lambda
^{2}\left\langle \Delta BX^{\dagger }X\Delta B\right\rangle _{\Psi
}+\lambda \left\langle \Delta BX^{\dagger }\Delta A+\Delta AX\Delta
B\right\rangle _{\Psi }\geq 0
\end{equation}
therefore the condition is obtained:
\begin{equation}
X^{\dagger }X=1
\end{equation}
The expression is positive for every value of $\lambda $ when:
\begin{equation}
\left\langle \Delta A^{2}\right\rangle _{\Psi }\left\langle \Delta
B^{2}\right\rangle _{\Psi }\geq \frac{1}{4}\left\langle
K\right\rangle _{\Psi }^{2}
\end{equation}
where
\begin{equation}
K=\Delta AX\Delta B+\Delta BX^{\dagger }\Delta A
\end{equation}
If a complex structure $J$ is available, we may choose
\begin{equation}
X:=\cos \theta +J\sin \theta \Longrightarrow X^{\dagger }X=1
\end{equation}
and look for the maximum of $\left\langle K(\theta )\right\rangle
_{\Psi }^{2}.$

We have
\begin{equation}
\left\langle K(\theta )\right\rangle _{\Psi }=\left\langle [\Delta
A,\Delta B]_{+}\right\rangle _{\Psi }\cos \theta +\left\langle
J[A,B]\right\rangle _{\Psi }\sin \theta =:\alpha \cos \theta +\beta
\sin \theta
\end{equation}
Then
\begin{equation}
\frac{d}{d\theta }\left\langle K(\theta )\right\rangle _{\Psi
}^{2}=0\Longrightarrow \tan \theta _{\max }=\frac{\beta }{\alpha }%
\Longrightarrow \left\langle K_{\max }\right\rangle _{\Psi
}^{2}=\alpha ^{2}+\beta ^{2}
\end{equation}
so that eventually
\begin{equation}
\left\langle \Delta A^{2}\right\rangle _{\Psi }\left\langle \Delta
B^{2}\right\rangle _{\Psi }\geq \frac{1}{4}\left\langle [\Delta
A,\Delta B]_{+}\right\rangle _{\Psi }^{2}+\frac{1}{4}\left\langle
J[A,B]\right\rangle _{\Psi }^{2}
\end{equation}
This general form of the indetermination relation, first discovered
by Schroedinger \cite{Schroe} and Robertson \cite{Robertson} in the
30's, reduces to the weaker Heisenberg relation for uncorrelated
states (see also \cite{sud}).

 \textbf{Remark} The same expression
for the indetermination relation holds also on the
\textsl{complexified} (\emph{via }$J$) Hilbert space.

\bigskip

\section{The Schroedinger picture}

Before we have discussed the need of a complex structure $J$ in
Quantum Mechanics. In this section we discuss the relevant geometric
structures which appear in Quantum Mechanics and the relations among
them.

To avoid technicalities, for the time being, while we deal with
general aspects, we shall confine ourselves to finite dimensional
carrier spaces; later, to consider the transition to Classical
Mechanics, we will deal with operators acting on infinite
dimensional Hilbert spaces.

We consider a vector space $V$ equipped with a complex structure
$J$, that is a (1-1)-tensor with the property $J^{2}=-1$. As $V$ is
a vector space its linear structure determines -and is determined
by- a vector field $\Delta $, the infinitesimal generator of
dilation (often it is also called the Liouville vector field or the
Euler operator).

If we introduce Cartesian coordinates for $V$, say $\{x^{j}\}$, we
have
\begin{equation}
\Delta =x^{j}\frac{\partial }{\partial x^{j}}
\end{equation}
while for the complex structure we find
\begin{equation}
J=J_{k}^{j}dx^{k}\otimes \frac{\partial }{\partial x^{j}}
\end{equation}
with the property

\begin{equation}
J_{k}^{j}J_{m}^{k}=-\delta _{m}^{j}.
\end{equation}

The existence of $J$ on the vector space $V$ implies that $\dim V$
is even, say $2n.$ We denote by $g$ the metric tensor (Euclidean
metric tensor on $V$) defined by
\begin{equation}
g=g_{jk}dx^{j}\otimes dx^{k}
\end{equation}
with $\det \left\| g_{jk}\right\| \neq 0$ and \ $g_{jk}x^{j}x^{k}$ a
positive-definite quadratic function.

The admissibility condition of $g$ and $J$ is stated \cite{Morandi}
by requiring that
\begin{equation}
g(Jv,Jw)=g(v,w),  \label{finJ}
\end{equation}
\begin{equation}
g(Jv,w)+g(v,Jw)=0  \label{infinJ}
\end{equation}
i.e. $J$ is an orthogonal transformation both at the finite level
(Eq.(\ref {finJ})) and at the infinitesimal level
(Eq.(\ref{infinJ})).

Out of $g$ and $J$ we may construct a symplectic structure $\omega $
defined by
\begin{equation}
\omega (v,w)=g(Jv,w).  \label{admJ}
\end{equation}
It follows that $\omega $ is not degenerate and
\begin{equation}
\omega (v,w)=-\omega (w,v),
\end{equation}
that is $\omega $ is indeed a symplectic structure.

Having these four structures on $V$, we may consider the group of
diffeomorphisms of $V$ and identify subgroups by requiring that one
or more of the previous structures are preserved.

By requiring that $\varphi _{\ast }\Delta =\Delta $ we collapse the
infinite dimensional diffeomorphism group to the finite dimensional
general linear group $GL(V).$

The requirement $\varphi ^{\ast }\omega =\omega $ provides us with
the infinite dimensional subgroup of symplectic (canonical)
transformations. While $\varphi ^{\ast }g=g$ gives us the finite
dimensional rotation group
and $\varphi ^{\ast }J=J$ identifies the general linear group $GL(n,\Bbb{C}%
). $

By taking the intersection of any two linear subgroups associated
with any pair of admissible structures we get the subgroup of
unitary transformations with infinitesimal generators provided by
the antiHermitian operators with respect to the Hermitian structure
\begin{equation}
h(v,w):=g(v,w)+i\omega (v,w)=:<v|w>.  \label{innerprod}
\end{equation}

We recall that an Hermitian structure on $V,$ as a real vector
space, can be defined as a map:
\begin{equation}
h:V\times V\longrightarrow \Bbb{R}^{2}  \label{targhet}
\end{equation}
such that

\begin{equation}
h(v,w)\longrightarrow (g(v,w),\omega (v,w)).
\end{equation}
We can also induce a complex structure on $V$ by setting, for
$z=(\alpha
+i\beta )\in \Bbb{C}$ and $u\in V,$%
\begin{equation}
zu=(\alpha +i\beta )u=\alpha u+J(\beta u).
\end{equation}
To consider the product structure on functions and more specifically
on
quadratic functions, it is convenient to consider directly the target space $%
\Bbb{R}^{2},$ in Eq.(\ref{targhet}), as $\Bbb{C},$ the complex
numbers. In this way, if $V$ is considered as a real vector space,
\[
h(v,w)=g(v,w)+i\omega (v,w)
\]
is a complex valued quadratic function of real variables.

\textbf{Remarks}

i)\ Notice that if the admissibility condition of Eq.(\ref{finJ})
does not hold, we can always build a Hermitian structure out of a
given $g$ by substituting it with the symmetrized metric tensor:
\begin{equation}
g_{s}(.,.)=:\frac{1}{2}\{g(J.,J.)+g(.,.)\}.
\end{equation}
which will be positive and nondegenerate if $g$ is.

Quite similarly \cite{Abraham}, if the relation of  Eq.(\ref{admJ})
does not define a complex structure $J$, that is if $(g^{-1}\omega
)^{2}\neq -\Bbb{I}$ , then Riesz's theorem tells us that there
exists a nonsingular linear operator $A$ such that:
\begin{equation}
\omega (x,y)=g(Ax,y)
\end{equation}
and the antisymmetry of $\omega $ implies:
\begin{equation}
g(Ax,y)=-g(x,Ay)
\end{equation}
i.e. that $A$ is skew-hermitian: $A^{\dagger }=-A$, and: $-A^{2}>0$.
Let then $P$ be a (symmetric) nonnegative square root of $-A^{2}$.
$P$ will be
injective, so $P^{-1}$ will be well defined\footnote{%
In the infinite-dimensional case, it can be proved that $A$ is
bounded and
injective, and that $P$ is also injective and densely defined, so that $%
P^{-1}$ is well defined in the infinite-dimensional case as well.}.
We define then: $J=:AP^{-1}$ and:
\begin{equation}
g_{\omega }(.,.)=:g(P(.),.)
\end{equation}

Therefore:
\begin{equation}
\omega(x,y)=g(Ax,y)=g_{\omega}(Jx,y)
\end{equation}
and: $J^{\dagger}=-J,$ $J^{2}=-\Bbb{I}$ . The triple
$(g_{\omega},J,\omega)$ will be then an admissible triple,
Eq.(\ref{infinJ}) will hold true for $g_{\omega}$ and, moreover:
\begin{equation}
g_{\omega}(Jx,Jy)=g(Ax,Jy)=-g(AJx,y)=g_{\omega}(x,y)
\end{equation}
and Eq.(\ref{finJ}) will be satisfied as well.

ii) \ The adjoint $A^{\dagger}$ of any linear operator $A$ with
respect to a metric tensor $g$ is defined by the standard relation:
\begin{equation}
g(A^{\dagger}x,y)=:g(x,Ay)\label{acroce}
\end{equation}
and we can read Eq.(\ref{finJ}) as saying that the complex structure
$J$ is skew-adjoint with respect to the metric tensor $g.$

Although it may seem elementary, it is worth stressing here that,
despite the fact that we are working in a real vector space, the
adjoint of $A$ does \textbf{not} coincide with the transpose $A^{T}$
for a general $g$. Indeed, spelling out explicitly Eq.(\ref{acroce})
in terms of matrices leads to:
\begin{equation}
A^{\dagger }=g^{-1}A^{T}g
\end{equation}
and therefore, even for real matrices: $A^{\dagger }=A^{T}$ only if
the metric is standard Euclidean one and, in general, symmetric
matrices need not be self-adjoint.

\bigskip

To completely turn entities depending on the linear structure on the
space of states into tensorial objects, we notice that with every
matrix $A\equiv \left\| A_{k}^{j}\right\| \in
\mathfrak{gl}(2n,\Bbb{R})$ we can associate both a $(1-1)-$tensor
\begin{equation}
T_{A}=A_{k}^{j}dx^{k}\otimes \frac{\partial }{\partial x^{j}}
\end{equation}
and a linear vector field
\begin{equation}
X_{A}=A_{k}^{j}x^{k}\frac{\partial }{\partial x^{j}}.
\end{equation}
The two are connected by
\begin{equation}
T_{A}(\Delta )=X_{A}
\end{equation}
and are both homogeneous of degree zero, i.e.
\begin{equation}
L_{\Delta }X_{A}=L_{\Delta }T_{A}=0.
\end{equation}

The correspondence $A\longrightarrow T_{A}$ is a full associative
algebra
and a corresponding Lie algebra isomorphism. The correspondence $%
A\longrightarrow X_{A}$ is instead only a Lie algebra
(anti)isomorphism, that is
\begin{equation}
T_{A}\circ T_{B}=T_{AB}
\end{equation}
while
\begin{equation}
\lbrack X_{A},X_{B}]=-X_{[A,B]}.
\end{equation}
Moreover, for any $A,B\in \mathfrak{gl}(2n,\Bbb{R})$ :
\begin{equation}
L_{X_{A}}T_{B}=-X_{[A,B]}.
\end{equation}
Out of the Liouville vector field $\Delta $ and the metric tensor
$g$ , we can construct the quadratic function
\begin{equation}
\mathfrak{g}\Bbb{=}\frac{1}{2}g(\Delta ,\Delta )  \label{met}
\end{equation}
along with the associated Hamiltonian vector field $\Gamma $ via the
\begin{equation}
i_{\Gamma }\omega =-d\mathfrak{g.}  \label{gamma}
\end{equation}
It is possible to show that
\begin{equation}
\Gamma =J(\Delta )  \label{jdelta}
\end{equation}
or $J(\Gamma )=-\Delta $.

The vector field $\Gamma $ preserves all three structures $g,\omega
$ and $J$ . Thus the vector field $\Gamma $ will be a generator of
the unitary group and may be associated with a Schroedinger-type
equation
\begin{equation}
i\hbar \frac{d}{dt}\Psi =L_{\Gamma }\Psi =H\Psi ,
\end{equation}
where we have used the more familiar notation $\Psi \in V$ instead
of $\ u\in V$ .

More generally, starting from Eqs.(\ref{met}), (\ref{gamma}) and
(\ref{jdelta}), we can construct quadratic functions associated with
any linear vector field $X_{A}$ by defining
\begin{equation}
f_{A}\Bbb{=}\frac{1}{2}g(\Delta ,X_{A})\ .  \label{quad}
\end{equation}
Then again we have
\begin{equation}
i_{Y_{A}}\omega =-df_{A}
\end{equation}
where now
\begin{equation}
Y_{A}=J(X_{A})\   \label{grad-ham}
\end{equation}
whenever the matrix $A$ satisfies the condition of the following
Eq.(\ref{omegaA+Aomega}).

In general, any vector field $X_{A}$ will define the equations of
motion of a Quantum systems if
\begin{equation}
L_{X_{A}}\omega =0\ \ ;\ L_{X_{A}}g=0
\end{equation}
and, \emph{a fortiori , \ }$L_{X_{A}}J=0$.

This implies that the matrix $A$, representative of the vector
field, satisfies the relations
\begin{equation}
\omega A=(\omega A)^{tr}\Longleftrightarrow \omega A+A^{tr}\omega =0
\label{omegaA+Aomega}
\end{equation}
while the condition $L_{X_{A}}g=0$ implies that
\begin{equation}
gA+(gA)^{tr}=gA+A^{tr}g=0,
\end{equation}
where, with abuse of notation, we have used $\omega $ and $g$ to
represent a matrix associated with the corresponding tensors.

By using the vector fields $\Delta $ and $J(\Delta )$, which are an
involutive distribution, i.e. in particular they commute, we get out of $%
V-\{0\}$ the complex projective space associated with the Hilbert
space
structure on $V$ defined by the Hermitian form $h$ (see Eq.(\ref{innerprod}%
)), which will be denoted by $\Bbb{P}(V)$.

By construction, the unitary group acting on $V$ in terms of linear
transformation will induce an action on $\Bbb{P}(V)$ which turns out
to be a transitive action.

The first outcome of our description in terms of tensors is that now
instead of linear operators on $\Bbb{P}(V)$ we can use vector fields
on $\Bbb{P}(V)$ which are autonomously defined without making
recourse to the underlined vector space $V.$ The main point here is
that $\Bbb{P}(V)$ \ is not a linear space therefore it makes no
sense to consider on it linear operators, one has to induce an
action from linear operators acting on $V$.

The use of vector fields allows to bypass the construction of the action in terms of $%
V. $

\section{The momentum map associated with the symplectic action of the
unitary group}

We have remarked that the linear action of the unitary group on $V$
is provided by the intersection of the symplectic group and the
rotation group.

Because the group acts symplectically, there will be an associated
momentum map from the vector space to the dual of the Lie algebra,
say
\begin{equation}
\mu :V\longrightarrow \mathfrak{u}^{\ast }(n)
\end{equation}
in explicit terms we have
\begin{equation}
\mu (v)=i\ |v><v|\ ,\ \ \ v\in V
\end{equation}
where we have used the bra and ket notation of Dirac, and identified $%
\mathfrak{u}^{\ast }(n)$ with $\mathfrak{u}(n)$ through a trace
operation. By using our previous structures, it would be
\begin{equation}
\mu (v)=i \ v \ h(v,.)
\end{equation}
which acts on $\mathfrak{u}(n)$ \ in the following way:
\begin{equation}
\mu (v)(A)=i<v|A|v>
\end{equation}
for any $A\in \mathfrak{u}(n).$

To have a momentum map equivariant with respect to the action of
$\Delta $ and $J(\Delta )$ \ we may set
\begin{equation}
\widetilde{\mu }(v)=i \ \frac{|v><v|}{<v|v>}
\end{equation}
or
\begin{equation}
\widetilde{\mu }(v)=i \ \frac{vh(v,.)}{h(v,v)}\ .
\end{equation}
This map represents the momentum map associated with the symplectic
action of $U(n)$ on the complex projective space $\Bbb{P}(V).$

It is now clear  that all vector fields associated with Hermitian
operators
on $V$ will pass to the quotient $\Bbb{P}(V)$ because they commute with $%
\Delta $ and preserve $J$ .

With any Hermitian operator $A$ we associate $X_{A}$ which projects onto $%
\Bbb{P}(V)$, giving a symplectic vector field with generating
function
\begin{equation}
\tilde{f}_{A}(\Psi )=\frac{<\Psi |A\Psi >}{<\Psi |\Psi >}
\end{equation}
that is
\begin{equation}
\tilde{f}_{A}(\Psi )=\frac{h(\Psi ,A\Psi )}{h(\Psi ,\Psi )}\ .
\end{equation}
We observe that functions here defined may be expressed through the
quadratic functions defined by Eq.(\ref{quad}) of the previous
Section: they are nothing but the
quadratic functions projected onto $\Bbb{P}(V)$%
\begin{equation}
\tilde{f}_{A}(\Psi )=\frac{<\Psi |A\Psi >}{<\Psi |\Psi
>}=\frac{g(\Delta ,X_{A})}{g(\Delta ,\Delta )}
\end{equation}
It is now possible to formulate the equations of motion in terms of
this ''mean values'' of the Hermitian operators.

\section{The Ehrenfest picture}

The Ehrenfest picture \cite{EMS} of Quantum Mechanics originates
from the Ehrenfest theorem. We may formulate it within our approach
in the following way.

We consider the Lie algebra of the unitary group realized in terms
of skew-Hermitian operators on $V$ and we define a map
\begin{equation}
E:\mathfrak{u}(n)\times V\longrightarrow \Bbb{R\ \ \ ;\ \ }(\imath
A,v)\longrightarrow \frac{h(v,Av)}{h(v,v)}=\tilde{f}_{A}(v).
\end{equation}
This evaluation map has the property that
\begin{equation}
\widetilde{E}:V\longrightarrow Lin(\mathfrak{u}(n),\Bbb{R)\equiv }%
\mathfrak{u}^{\ast }(n)
\end{equation}
coincides with the momentum map associated with the symplectic
action of the unitary group acting on $V.$

Our aim now is to realize the algebra of linear operators on $V$ in
the Ehrenfest picture, that is by means of quadratic functions. We
observe that the definition of Eq.(\ref{quad}) in Sec.(3)is not
suitable to this end, because the
real scalar product $g$ annihilates the skew-symmetric part of any operator $%
A.$ Therefore, we have to use the Hermitian form $h$ defined by
Eq.(\ref {innerprod}) of Sec.(3) and deal with complex-valued
functions of real variables:
\begin{equation}
f_{A}(x)=\frac{1}{2}h(Ax,x)\ ;\ x\in V.
\end{equation}
This definition reduces to the previous one and yields real
functions for Hermitian operators, while associates imaginary
functions to skew-hermitian operators, so that it realizes a one to
one correspondence between operators
and complex-valued quadratic functions. We notice that neither $g$ nor $%
\omega $ separately can be used to recover the full associative
algebra: in fact $\omega $ allows to recover the Lie algebra
structure of the Hermitian operators, but it is not able to recover
the symmetrized product on Hermitian operators, i.e. the Jordan
algebra structure \cite{Jordan} existing on the space of Hermitian
operators. To recover this product we have to use the metric tensor
$g$. It is only the use of the entire admissible triple $(g,\omega
,J)$ or the equivalent Hermitian structure $h$ that allows to give a
complete description of the full algebra of operators in the
Ehrenfest scheme.

We now introduce two brackets on quadratic functions corresponding
respectively to Lie and Jordan operator algebras and from them
recover the associative operator product.\cite{Cirelli} In fact, it
is well known that on the
symplectic vector space $(V,\omega )$ we can define a Poisson tensor $%
\Lambda $ which is the ''inverse'' of $\ \omega $: in coordinates we
have
\begin{equation}
\Lambda =\Lambda ^{jk}\frac{\partial }{\partial x^{j}}\wedge
\frac{\partial }{\partial x^{k}}
\end{equation}
with $\Lambda ^{jk}\omega _{km}=\delta _{m}^{j}$ .

The associated Poisson Bracket is given by
\begin{equation}
\{f_{1},f_{2}\}:=\Lambda (df_{1},df_{2})=\Lambda ^{jk}\frac{\partial f_{1}}{%
\partial x^{j}}\frac{\partial f_{2}}{\partial x^{k}}.
\end{equation}
On quadratic functions we have the following result:
\begin{equation}
\{f_{A},f_{B}\}(x)=f_{J[A, B]}(x)=-if_{[A, B]}(x)\ .
\end{equation}

In fact, supposing $\ A,B$ Hermitian operators, in coordinates
\begin{equation}
f_{A}=\frac{1}{2}g_{ml}A_{n}^{l}x^{n}x^{m}\ ;\ f_{B}=\frac{1}{2}%
g_{sp}B_{t}^{p}x^{t}x^{s}.
\end{equation}
Then, bearing in mind that
\begin{equation}
\Lambda ^{kj}g_{jm}=-J_{m}^{k},
\end{equation}
we have:
\[
\Lambda ^{jk}\frac{\partial f_{A}}{\partial x^{j}}\frac{\partial f_{B}}{%
\partial x^{k}}=\Lambda ^{jk}(Ax)_{j}(Bx)_{k}=\Lambda
^{jk}g_{mj}g_{nk}(Ax)^{m}(Bx)^{n}=
\]
\[
=-\frac{1}{2}g_{mj}J_{n}^{j}(Bx)^{n}(Ax)^{m}+\frac{1}{2}%
g_{nk}J_{m}^{k}(Ax)^{m}(Bx)^{n}=
\]
\begin{equation}
=\frac{1}{2}\omega
([A,B]x,x)=-\frac{i}{2}h([A,B]x,x)=-if_{[A,B]}(x).
\end{equation}
Moreover

\begin{equation}
f_{[A,B]}(x)\ =\frac{1}{2}h([A,B]x,x)=\frac{i}{2}\omega
([A,B]x,x)=\frac{i}{2}g(J[A,B]x,x)=if_{J[A,B]}(x)
\end{equation}
The proof is analogous in the cases when one or both operators \ $A$
and $B$ are skew-symmetric, due to the relation $\omega =gJ$.

By using in a similar way the inverse of the metric tensor $%
g=g_{ik}dx^{j}\otimes dx^{k},$ say
\begin{equation}
G=G^{jk}\frac{\partial }{\partial x^{j}}\otimes \frac{\partial
}{\partial x^{k}}
\end{equation}
with $\ G^{jk}g_{km}=\delta _{m}^{j}$ , we may define a new Bracket
(we shall call it the Riemann-Jordan Bracket) on functions as:
\begin{equation}
(f_{1},f_{2}):=G(df_{1},df_{2})=G^{jk}\frac{\partial f_{1}}{\partial x^{j}}%
\frac{\partial f_{2}}{\partial x^{k}}.
\end{equation}
By using quadratic functions we get the relation
\begin{equation}
(f_{A},f_{B})(x)=f_{(AB+BA)}(x).
\end{equation}

In fact, assuming again $A,B$ Hermitian operators, and bearing in
mind that $G^{kj}g_{jm}=\delta _{m}^{k}$, we have in coordinates
\[
G^{jk}\frac{\partial f_{A}}{\partial x^{j}}\frac{\partial
f_{B}}{\partial
x^{k}}=G^{jk}(Ax)_{j}(Bx)_{k}=G^{jk}g_{mj}g_{nk}(Ax)^{m}(Bx)^{n}=
\]
\[
=\frac{1}{2}g_{nk}\delta
_{m}^{k}(Ax)^{m}(Bx)^{n}+\frac{1}{2}g_{mj}\delta
_{n}^{j}(Bx)^{n}(Ax)^{m}=\frac{1}{2}g([A,B]_{+}x,x)=
\]
\begin{equation}
=\frac{1}{2}h([A,B]_{+}x,x)=f_{(AB+BA)}(x).
\end{equation}

Finally, we introduce an associative product on quadratic functions
as
\begin{equation}
(f_{A}\ast f_{B})(x):=f_{AB}(x)\ \ ,
\end{equation}
so that
\begin{equation}
(f_{A}\ast f_{B})(x)+(f_{B}\ast
f_{A})(x)=f_{(AB+BA)}(x)=(f_{A},f_{B})(x)
\end{equation}
and
\begin{equation}
(f_{A}\ast f_{B})(x)-(f_{B}\ast
f_{A})(x)=f_{(AB-BA)}(x)=i\{f_{A},f_{B}\}(x)\ ,
\end{equation}
so we obtain the following result:
\begin{equation}
(f_{A}\ast f_{B})(x)=\frac{1}{2}(f_{A},f_{B})(x)+\frac{i}{2}%
\{f_{A},f_{B}\}(x),
\end{equation}
which is the analog of the operator decomposition $AB=\frac{1}{2}[A,B]_{+}+%
\frac{1}{2}[A,B]$.

It is also apparent that the following relation which connects
gradient and Hamiltonian fields holds (see Eq.(\ref{grad-ham}) of
Sec.(3)):
\begin{equation}
grad\ f_{A}:=Gdf_{A}=-J\Lambda df_{A}.
\end{equation}

Equations of motion in this picture have the Classical-like form
\begin{equation}
\hbar \frac{d}{dt}f_{A}=\{f_{H},f_{A}\}.
\end{equation}
Within the one-to-one correspondence between operators and quadratic
functions, we recover the Heisenberg equations of motion in
Hamiltonian form.

We conclude this section by observing that the $\ast -$ product is
reminiscent of the Weyl-Wigner approach and the Moyal Bracket. It
shows also that a full isomorphism of algebras between quadratic
functions and operators requires the introduction of a non-local
product.

Summarizing, we have replaced the initial description provided by
Dirac in terms of Hilbert spaces and operators acting on them with a
description on the differentiable manifold of pure states (the
complex projective space) in terms of vector fields and functions.
It is clear that if we do not have a linear structure we have no
meaning of quadratic functions, therefore the ''image'' of quadratic
functions on $V$ shall be identified autonomously.

This identification is achieved by requiring that Hamiltonian vector
fields associated with them should be Killing vectors with respect
to the metric structure on $\Bbb{P}(V)$ arising from the K\"{a}hler
structure. This ends up being the infinitesimal version of Wigner's
theorem.\cite{Wigner}

\section{The Classical Limit}

To consider the Quantum-Classical transition, the best way is to
consider a description which uses the same carries space for both
Classical and Quantum descriptions, i.e. smooth functions on the
phase space. The Classical
picture uses the point-wise product, while the Quantum picture uses the $%
\ast -$product. In particular we shall restrict to a phase-space
which is a symplectic vector space and the $\ast $-product is the
Moyal product.

Given a symplectic vector space $(E,\omega )$, a Weyl system is
defined to be a strongly continuous map from $E$ to unitary
transformations on some Hilbert space $\Bbb{H}$ :
\begin{equation}
W:E\rightarrow U(\Bbb{H})
\end{equation}
with
\begin{equation}
W(e_{1})W(e_{2})W^{\dagger }(e_{1})W^{\dagger }(e_{2})=e^{^{\frac{{\Large i}%
}{{\Large \hbar }}{\Large \omega (e}_{1}{\Large ,e}_{2}{\Large
)}}}\Bbb{I},\label{weyl}
\end{equation}
with $\Bbb{I}$ the identity operator.

Consider a Lagrangian subspace $L$ and an associated isomorphism
\begin{equation}
E\rightleftharpoons L\oplus L^{\ast }=T^{\ast }L\ \ .
\end{equation}
On $L$\ we consider square integrable functions with respect to a
Lebesgue measure on $L$, a measure invariant under translations. The
splitting of $E$ allows to define $e=(\alpha ,x)$ and set
\begin{equation}
W((0,x)\Psi )(y)=\Psi (x+y),
\end{equation}
\begin{equation}
W((\alpha ,0)\Psi )(y)=e^{{\Large i\alpha (y)}}\Psi (y),
\end{equation}
\begin{equation}
x,y\in L\ ,\alpha \in L^{\ast },\Psi \in \emph{L}^{2}(L,d^{n}y);
\end{equation}
it is obvious that $W(e)$ are unitary operators and moreover they
satisfy Weyl condition of Eq.(\ref{weyl}) with $\omega $ being the
canonical one on $T^{\ast }L$.

The strong continuity allows to use Stone's theorem to get
infinitesimal generators $R(e)$ such that
\begin{equation}
W(e)=e^{{\Large i}R(e)}{\Large \ \ \ }\forall e\in E\
\end{equation}
and $R(\lambda e)=\lambda R(e)$ for any $\lambda \in \Bbb{R}$.

When we select a complex structure on $E$
\begin{equation}
J:E\rightarrow E\ \ ,\ \ J^{2}=-1\ \ ,
\end{equation}
we may define ''creation'' and ''annihilation'' operators by setting
\begin{equation}
a(e)=\frac{1}{\sqrt{2}}(R(e)+iR(Je)),
\end{equation}
\
\begin{equation}
a^{\dagger }(e)=\frac{1}{\sqrt{2}}(R(e)-iR(Je)).
\end{equation}
By using this complex structure on $E${\Large \ }we may construct an
inner product on $E$ as
\begin{equation}
\left\langle e_{1},e_{2}\right\rangle =\omega (Je_{1},e_{2})-i\omega
(e_{1},e_{2}),
\end{equation}
therefore creation and annihilation operators are associated with a
K\"{a}hler structure on $E.$

The Weyl map can be extended to functions on $T^{\ast }L\rightleftharpoons E$%
; indeed we first define the symplectic Fourier transform of $\hat{f}\in $ $%
\mathfrak{F}(E)$
\begin{equation}
f(q,p)=\frac{1}{2\pi \hbar }\int \hat{f}(\alpha ,x)e^{\frac{i}{\hbar }%
(\alpha q-xp)}d\alpha dx
\end{equation}
and then associate with it
the operator $\widehat{A}_{f}$ defined by
\begin{equation}
\widehat{A}_{f}=\frac{1}{2\pi \hbar }\int \hat{f}(\alpha ,x)e^{\frac{i}{%
\hbar }(\alpha \widehat{Q}-x\widehat{P})}d\alpha dx.
\end{equation}
\emph{Vice versa}, with any operator $A$ acting on $\Bbb{H}$ we
associate a function $f_{A}$ on the symplectic space $E$ by setting%
\begin{equation}
f_{A}(e):=TrAW(e)\ ;
\end{equation}
this map is called the Wigner map and, \emph{via} a symplectic
Fourier transform, is the inverse of the Weyl map. A new product of
functions may be introduced on $\mathfrak{F}(E)$ by setting
\begin{equation}
\left( f_{A}\star f_{B}\right) (e):=TrABW(e)=f_{AB}(e).
\end{equation}
We thus find that a symplectic structures on $E$ give rise to an
associative product on $\mathfrak{F}(E),$ which is not commutative.

The dynamics on $\mathfrak{F}(E)$ can be written in terms of this
non-commutative product as
\begin{equation}
i\hbar \frac{df_{A}}{dt}=f_{H}\star f_{A}-f_{A}\star f_{H}\ \ \ .
\end{equation}
Having clarified the setting, let us consider a very simple
situation to actually carry on the limiting procedure.

Without any loss of generality we may limit to consider a
bidimensional symplectic vector space
\begin{equation}
\Bbb{R}^{2}=\Bbb{R\oplus R};\ \omega =dx\wedge dy.
\end{equation}

The map $\hat{A}$ $\mapsto f_{A}$ given by $f_{A}(x,y)=Tr
\hat{A}W(x,y)$ induces a Moyal product
\begin{equation}
(f_{A}\star f_{B})(x,y):=Tr\hat{A}\hat{B}W(x,y)\ \ .
\end{equation}
By using the specific form of $W(x,y)$ it is possible to provide an
explicit form in terms of bidifferential operators. We have
\begin{equation}
(f_{A}\star f_{B})(x,y)=f_{A}(x,y)\,\exp \left[ i\frac{\hbar
}{2}\left(
\stackrel{\leftarrow }{\frac{\partial }{\partial x}}\stackrel{\rightarrow }{%
\frac{\partial }{\partial y}}-\stackrel{\leftarrow }{\frac{\partial }{%
\partial y}}\stackrel{\rightarrow }{\frac{\partial }{\partial x}}\right) %
\right] \,f_{B}(x,y)
\end{equation}
where a standard notation for physicists has been used, i.e. $\stackrel{%
\leftarrow }{\partial _{x}}$ and $\stackrel{\rightarrow }{\partial
_{y}}$ mean that the operators act on the left or on the right,
respectively.

Alternatively, we may write
\begin{equation}
(f_{A}\star f_{B})(x,y)=\,\exp \left[ i\frac{\hbar
}{2}\mathfrak{L}\right] \,f_{A}(x^{\prime },y^{\prime })f_{B}(x,y)
\end{equation}
where the bidifferential operator $\mathfrak{L}$ has been introduced%
\begin{equation}
\mathfrak{L}:=\left[ \frac{\partial }{\partial x^{\prime }}\frac{\partial }{%
\partial y}-\frac{\partial }{\partial y^{\prime }}\frac{\partial }{\partial x%
}\right] _{x^{\prime }=x,y^{\prime }=y}
\end{equation}
The action of $\mathfrak{L}$ may be expressed in terms of the
Poisson Bracket of two functions $f,g\in \mathfrak{F}(\Bbb{R}^{2})$
as:
\begin{equation}
\mathfrak{L}fg=f_{x}g_{y}-f_{y}g_{x}=:\{f,g\}
\end{equation}
\begin{equation}
\mathfrak{L}^{2}fg=\mathfrak{L}
\{f,g\}=\{f_{x},g_{y}\}-\{f_{y},g_{x}%
\}=f_{xx}g_{yy}-2f_{xy}g_{yx}+f_{yy}g_{xx}
\end{equation}
and in general (all $\partial $'s commute among themselves)
\begin{equation}
\mathfrak{L}
^{n+1}fg=\mathfrak{L}^{n}\{f,g\}=\sum\limits_{k=0}^{n}\left(
\begin{array}{c}
n \\
k
\end{array}
\right) %
(-1)^{k}\{\partial _{x}^{n-k}\partial _{y}^{k}f,\partial
_{y}^{n-k}\partial _{x}^{k}g\}
\end{equation}
so that
\begin{equation}
(f_{A}\star f_{B})(x,y)=f_{A}(x,y)f_{B}(x,y)+i\frac{\hbar }{2}%
\{f_{A},f_{B}\}+\frac{1}{2}(i\frac{\hbar }{2})^{2}\mathfrak{L}%
^{2}f_{A}f_{B}+...
\end{equation}
from which we see that the Quantum corrections to the point-wise
Abelian product are expressed by means of the complex structure
times the Planck's constant and the Poisson Bracket of the two
functions and of all their derivatives.

Now, even or odd powers of $\mathfrak{L}$\ are respectively
symmetric or skewsymmetric with respect to the exchange of $f_{A}$
and $f_{B}$ , so that
the commutator of $f_{A}$ and $f_{B}$ contains only the odd powers of $%
\mathfrak{L}:$
\begin{equation}
f_{A}\star f_{B}-f_{B}\star f_{A}=2i\frac{\hbar }{2}[\{f_{A},f_{B}\}+\frac{1}{6}(i%
\frac{\hbar }{2})^{2}\mathfrak{L}^{3}f_{A}f_{B}+...]
\end{equation}
and the Quantum equation of motion of $f_{B}$ with respect to the
Hamiltonian $f_{H}$ reads
\begin{equation}
i\hbar \frac{d}{dt}f_{B}=f_{H}\star f_{B}-f_{B}\star f_{H}=i\hbar
\lbrack \{f_{H},f_{B}\}+O(\hbar ^{2})]
\end{equation}
and finally, dropping terms $O(\hbar ^{2}),$ we recover the
Classical equations of motion in the Hamiltonian form:
\begin{equation}
\frac{d}{dt}f_{B}^{\circ }=\{f_{H},f_{B}^{\circ }\}
\end{equation}
When the Hamiltonian $f_{H}$ is a quadratic polynomial, the $%
O(\hbar ^{2})$ terms vanish, so this result is exact and we recover
the Ehrenfest's theorem.

This ''appropriate Classical limit'' shows very clearly that while
in Quantum Mechanics the associative and non commutative product
among functions determines all the various structures, i.e. the
complex structure and the Poisson bracket, because the Lie product
and therefore the Jordan product are uniquely determined by the
operator-product \cite{Gra-Ma}, in the Classical picture the
point-wise product does not fix a unique ''compatible Poisson
bracket''. Indeed all possible Poisson tensors on phase space define
derivations for the point-wise product on functions on the phase
space.

To exhibit more clearly this aspect, i.e. the role of the symmetric
tensor in the definition of the $\ast -$product we use a slightly
modified product with respect to the one given by Moyal. This
generalization arises from the fact that one may give a different,
more general, definition of Weyl systems where ''different ordering
procedures'' are considered instead of the fully symmetrized one
provided by the Weyl prescription.

It is the following one \cite{Aniello}
\begin{equation}
W(e_{1})W(e_{2})W^{\dagger }(e_{1})W^{\dagger }(e_{2})=e^{^{-g{\Large (e}_{1}%
{\Large ,e}_{2}{\Large )+i\omega (e}_{1}{\Large ,e}_{2}{\Large )}}}\Bbb{I}%
\ .
\end{equation}
In analogy with this modification we may consider a $\ast -$product
given by
\begin{equation}
(f_{A}\star f_{B})(x,y)=\,\exp \frac{\lambda }{2}\left[ \mathfrak{G}\frak{+}i%
\mathfrak{L}\right] \,f_{A}(x^{\prime },y^{\prime })f_{B}(x,y)
\end{equation}
where the bidifferential operator $\frak{G}$ has been introduced:
\begin{equation}
\mathfrak{G}:=\left[ \frac{\partial }{\partial x^{\prime }}\frac{\partial }{%
\partial x}+\frac{\partial }{\partial y^{\prime }}\frac{\partial }{\partial y%
}\right] _{x^{\prime }=x,y^{\prime }=y}.
\end{equation}
and we have used a dimensionless deformation parameter $\lambda $
instead of $\hbar .$

Now we get
\begin{equation}
(f_{A}\star f_{B})(x,y)=f_{A}(x,y)f_{B}(x,y)+\frac{\lambda
}{2}(\partial
_{x}f_{A}\partial _{x}f_{B}+\partial _{y}f_{A}\partial _{y}f_{B})+i\frac{%
\lambda }{2}\{f_{A},f_{B}\}+O(\lambda ^{2}).
\end{equation}

We consider the Classical limit of the Lie product and the Classical
limit of the Jordan product to find
\[
\frac{1}{2}(f_{A}\star f_{B}-f_{B}\star f_{A})(x,y)=i\frac{\lambda }{2}%
\{f_{A},f_{B}\}+O(\lambda ^{2}),
\]
\begin{equation}
\frac{1}{2}(f_{A}\star f_{B}+f_{B}\star f_{A})(x,y)=f_{A}(x,y)f_{B}(x,y)+%
\frac{\lambda }{2}(\partial _{x}f_{A}\partial _{x}f_{B}+\partial
_{y}f_{A}\partial _{y}f_{B})+O(\lambda ^{2})
\end{equation}
respectively.

We see immediately that if the product should reduce to the
commutative point-wise product the symmetric bidifferential operator
cannot be there.

\section{Conclusions and outlook}

Let us summarize our line of argumentation. In Quantum Mechanics the
product of operators (observables) uniquely defines the Lie product
compatible with it (i.e. such that it defines derivations of the
given product).

If we associate operators with functions (a necessary replacement if
we want to deal with physical pure states, points of the complex
projective space), the product among operators induces a $\star
-$product on functions. In the Weyl-Wigner approach this product can
be expressed by means of bidifferential operators (generated by a
Poisson tensor in the case of the Weyl ordering and by a Poisson and
a metric tensors in the case of normal or anti-normal ordering).

In the Classical limit, the product gives rise to a point-wise
product which does not privilege any specific Poisson Bracket.
However, a Poisson Bracket may be selected by considering the
Classical limit of the Lie product. A similar Classical limit for
the Jordan product would anyway start with the point-wise term and
therefore would not give rise naturally to a symmetric tensor on
phase-space.

Our argument has been presented without privileging any dynamical
system. It is however true that if we consider special systems like
the Harmonic Oscillator, by considering the limit of operators
realized in terms of creation and annihilation operators, we would
get a Classical phase-space described in terms of complex
coordinates. However, these would not exist for generic Hamiltonian
systems.

As a further comment, we may add that the Classical limit may be
also considered in the Schroedinger picture.

Here the the Classical limit would require first that the operators
are realized as differential operators on some ``configuration
space'', acting on square integrable functions defined on the same
configuration space.

The Classical limit would correspond to the replacement of the
differential operators with their symbols \cite{EMS} and the complex
structure would appear only at the level of the so called ``Quantum
potential'' having $\ \hbar ^{2}$ as coefficient. Moreover, out of
the coupled equations for the real and the imaginary part of the
wave function, we would obtain only one equation (for $\hbar
^{2}\rightarrow 0$) represented by the Hamilton-Jacobi equation. A
complete solution for this equation would allow for the construction
of a solution for the other one (the transport equation).\cite{Van
Vleck}

Therefore, having obtained uncoupled equations in the Classical
limit, the complex structure is not required to couple them as in
the Quantum situation.

We shall further elaborate on these points somewhere else.


\begin{thebibliography}{99}
\bibitem{1}  P.A.M. Dirac,\emph{\ The Principles of Quantum Mechanics,}
4th Edition (Pergamon, Oxford, 1958).

\bibitem{Weil} A. Weil,\emph{\ Introduction \`{a} l'Etudes Vari\'{e}t\'{e}s
K\"{a}hleriennes} (Hermann, Paris, 1958).

\bibitem{Carinena}  J.F. Cari\~{n}ena, J. Grabowski and G. Marmo, \emph{%
Lie-Scheffers Systems: A geometric Approach } (Bibliopolis, Napoli,
2000).

\bibitem{Sudarshan} V.I.Mank'o, G. Marmo, E.C.G. Sudarshan, F.
Zaccaria, Interference and entanglement: an intrinsic approach,
\emph{ J. Phys. A: Math. and Gen.} \textbf{35} (2002), 7137-7157.

\bibitem{PanchaMukunda} S. Pancharatnam, Generalized theory of
interference, and its applications, \emph{Proc. Ind. Acad. Sci.}
\textbf{A 44} (1956), 247-262.

\bibitem{Reichenbach} H. Reichenbach \emph{Philosophic Foundations of
Quantum Mechanics } (University of California Press , Berkeley and
Los Angeles, 1944).

\bibitem{Stuec}  E.C.G. Stueckelberg, Quantum Theory in Real Hilbert
Space, \emph{Helv Phys Acta} \textbf{33} (1960), 727-751;\\E.C.G.
Stueckelberg and M. Guenin, Quantum Theory in Real Hilbert Space II
(Addenda and Errata), \emph{Helv Phys Acta} \textbf{34} (1961), 621.

\bibitem{Mackey} G. W. Mackey, \emph{Mathematical Foundations of Quantum Mechanics}
(Benjamin, New York, 1963);\\ G. W. Mackey, \emph{Induced
Representations of Groups and Quantum Mechanics} (Benjamin, New
York, 1968).

\bibitem{Cantoni} V. Cantoni, Generalized ''Transition Probability'', \emph{Commun. Math. Phys. }%
\textbf{44} (1975), 125-128;\\
V. Cantoni, The Riemannian Structure on the States of Quantum-like
Systems, \emph{Commun. Math. Phys. }\textbf{56} (1977), 189-193;\\
V. Cantoni,Intrinsic geometry of the quantum-mechanical ''phase
space'', hamiltonian systems and Correspondence Principle,\emph{
Lincei-Rend. Sc. fis. mat. e nat.} \textbf{LXII} (1977), 628-636.

\bibitem{Cirelli}  R. Cirelli, A. Mani\`{a} and L. Pizzocchero, Quantum
Mechanics as an infinite-dimensional Hamiltonian systems with
uncertainty structure: Part I, \emph{J. Math. Phys. }\textbf{31
}(1990), 2891-2897;\ \ R. Cirelli, A. Mani\`{a} and L. Pizzocchero,
Quantum Mechanics as an infinite-dimensional Hamiltonian systems
with uncertainty structure: Part II, \emph{J. Math. Phys.
}\textbf{31 }(1990), 2898-2903.

\bibitem{Weinberg}  S. Weinberg, Testing quantum Mechanics, \emph{ Ann.
Phys.} \textbf{194} (1989), 336-386;\\
S. Weinberg, Precision tests of Quantum
Mechanics,\emph{Phys. Rev. Lett.} \textbf{62} (1989), 485-488;\\
S. Weinberg, Weinberg replies,\emph{Phys. Rev. Let.} \textbf{63}
(1989), 1115.

\bibitem{Schroe}  E. Schroedinger, Zum Heisenbergshen Unsh\"{a}rfeprinzip,
\emph{Ber. Kgl. Akad. Wiss.}\textbf{296}(1930), 296-303.

\bibitem{Robertson}  H.P. Robertson, A general formulation of the
uncertainty principle and its classical interpretation, \emph{Phys. Rev.} \textbf{%
35} (1930), 667;\\
H.P. Robertson, An Indeterminacy Relation for Several Observables
and Its Classical Interpretation, \emph{Phys. Rev.} \textbf{%
46} (1934), 794-801.
\bibitem{sud} E.C.G. Sudarshan, C.B. Chiu, and G. Bhamati, Generalized uncertainty relations and
characteristic invariance for the multimode states, \emph{Phys. Rev.
A} \textbf{52} (1995), 43-54.

\bibitem{Morandi}  G. Marmo, G. Morandi, A. Simoni and F. Ventriglia, Alternative structures and
bi-Hamiltonian systems, \emph{%
J.Phys. A: Math.Gen.} \textbf{35} (2002), 8393-8406;\\
A. Canas da Silva, \emph {Lectures on Symplectic Geometry}
(Springer, Berlin, 2001).

\bibitem{Abraham}  R. Abraham and J. E. Marsden, \emph{Foundations of
Mechanics}, $2^{nd}$ Ed. (Addison-Wesley, Reading, 1985).

\bibitem{EMS}  G. Esposito, G. Marmo and G. Sudarshan, \emph{From Classical to
Quantum Mechanics} (Cambridge University Press, Cambridge, 2004).

\bibitem{Jordan}  P. Jordan, J. von Neumann, E. Wigner, On an algebraic
generalization of the quantum mechanical formalism, \emph{Ann.
Math.} \textbf{35 }(1934), 29-64.

\bibitem{Wigner}E. P. Wigner, \emph{ Group theory and its application to the quantum Mechanics of
atomic spectra} (Academic Press, New York-London, 1959)

\bibitem{Gra-Ma}  J. Grabowski, G. Marmo, \emph{Binary operations in classical and quantum Mechanics},
in \ J. Grabowski,
P. Urbanski (eds.),
\emph{Classical and Quantum integrabiblity}, Banach Center Publ. \textbf{59 }%
, 163-172 (2003).


\bibitem{Aniello}  P. Aniello, V.I. Mank'o, G. Marmo,S. Solimeno,F. Zaccaria
On the coherent states, displacement operators and
quasidistributions associated with deformed quantum oscillator,
\emph{Jour. of Optics B} \textbf{2}(2000), 718-725.


\bibitem{Van Vleck} V.A. Fock, \emph{Fundamentals of Quantum Mechanics} (MIR,
Moscow, 1978)



\end{thebibliography}
\end{document}